\documentclass{article}
\usepackage{amsmath}
\usepackage{amssymb}
\usepackage{amsfonts}
\usepackage{amscd}
\usepackage{latexsym}
\usepackage{revsymb}
\usepackage{graphicx}

\newtheorem{thm}{Theorem}

\newenvironment{pf}{\noindent{\bf Proof}\quad}{\hfill$\Box$\bigskip}

\newcommand{\ketbra}[1]{| #1 \rangle \langle #1 |}

\title{Yet another additivity conjecture}
\author{
$\mbox{\large{\bf Keiji Matsumoto}}^{1,2}$
\\
{\tt keiji@nii.ac.jp}
}

\begin{document}
\maketitle

\footnotetext[1]{
  ERATO Quantum Computation and Information Project, JST. \\
  Hongo White Building, 5-28-3 Hongo, Bunkyo-ku, Tokyo 113-0033, Japan.
}
\footnotetext[2]{
  Quantum Information Science Group, National Institute of Informatics. \\
  2-1-2 Hitotsubashi, Chiyoda-ku, Tokyo 101-8430, Japan.
}

\abstract{ 
In quantum information theory, there are several important open problems
which center around whether certain quantities are additive or not.
Especially, additivity conjecture about Holevo capacity and additivity/strong
superadditivity conjecture about entanglement of formation have been
attracting many researchers. It recently turned out that these are
equivalent to the additivity of the minimum output entropy of quantum
channels, which is mathematically simpler. This paper suggests yet
another additivity conjecture which is equivalent to those, and is
mathematically simple. This conjecture might be easier than other
conjectures to solve, for this can be proven for almost all the examples where
one of these conjectures are proven.  
}


\section{Introduction}

In quantum information theory, several open problems center
around whether certain quantities are additive or not. The additivity of
Holevo capacity is the oldest of those. 
As mathematical lemma to study
this conjecture, many researchers have been studying 
the additivity of the minimum output entropy of 
quantum channels. 

Also, the entanglement of formation (EoF) is conjectured to
be additive by many. 
For this quantity, the property called strong superadditivity is
conjectured \cite{VW}, too. Roughly speaking, this conjecture
insists that 
the sum of entanglement of
the subsystems should not be larger than entanglement of the whole
system.

As is proved in \cite{Shor:equivalence},  these additivity
conjectures are equivalent. Based on this result, now, 
many authors are working on additivity of the minimum output 
entropy of quantum channels, for its mathematical simplicity.

In this talk, we suggest yet another entanglement 
quantity, whose
strong superadditivity and additivity 
are equivalent to additivity of
the quantities mentioned above. 
The motivations are as follows. 
First, in existing proofs of additivity conjectures
for specific examples, they are 
essentially proving additivity of
this quantity. Second, this quantity seems 
at least as simple as the output minimum
entropy. Third, the author prefer entanglement 
measures rather than quantities about channels.

\section{Additivity conjectures}
Let $\rho $ be a bipartite state on 
$\mathcal{H}_{1}\otimes\mathcal{H}_{2}$. 
The \emph{entanglement of formation (EoF)} of $\rho $ is defined as 
\begin{equation}
E_{f}(\rho )
:=\min_{\{p_{i},\pi_i\}} \sum_i p_i E(\pi_i)
\label{eq:entf}
\end{equation}%
where $\{p_{i},\pi_{i}\}$ runs all over the ensembles of pure bipartite
states with $\sum_{i}p_{i}\pi _{i}=\rho $, and the \emph{(entropy of)
entanglement} for a pure bipartite state $\pi $ is defined as 
\begin{equation*}
E(\pi ):=S\left( \mathrm{tr}_{\mathcal{H}_{2}}\pi \right) 
=S\left( \mathrm{tr}_{\mathcal{H}_{1}}\pi \right) .
\end{equation*}%

Let $\rho $ be a state on $\mathcal{H}\otimes \mathcal{H}^{\prime }$, 
where $\mathcal{H}=\mathcal{H}_{1}\otimes \mathcal{H}_{2}$ 
and 
$\mathcal{H}^{\prime}=\mathcal{H}_{1}^{\prime }\otimes \mathcal{H}_{2}^{\prime }$. Then the
strong superadditivity \cite{VW} means that 
\begin{equation}
E_{f}(\rho )
\geq E_{f}(\mathrm{tr}_{\mathcal{H}^{\prime }}\rho )
+E_{f}(\mathrm{tr}_{\mathcal{H}}\rho ), 
\label{eq:entf:superadd}
\end{equation}%
where all entanglement of formation are understood with respect to the 
$1$--$2$--partition of the respective system. The "weaker version" 
of additivity conjecture of EoF states, 
\begin{equation}
E_{f}(\rho \otimes \rho ^{\prime })=E_{f}(\rho )+E_{f}(\rho ^{\prime }).
\label{entadd}
\end{equation}

Let $\Lambda $ be a CPTP map from $\mathcal{B(K)}$ to
${\cal B}({\cal H}_1)$. 
The \textit{minimum output entropy }is defined as,%
\begin{equation*}
S_{\rm min}\left( \Lambda \right) 
:=\min_{\rho \in \mathcal{S}\left( \mathcal{K}\right) }
S\left( \Lambda(\rho) \right) .
\end{equation*}
The additivity conjecture about this quantity means,
\begin{equation}
S_{\rm min}( \Lambda \otimes \Lambda^{\prime}) 
=S_{\rm min}(\Lambda ) 
+S_{\rm min}(\Lambda ^{\prime}) ,
\label{addminent}
\end{equation}

Shor\cite{Shor:equivalence} had proven that (\ref{eq:entf:superadd}), 
(\ref{entadd}), (\ref{addminent}), and additivity of the Holevo capacity are
equivalent with each other. In the proof, the correspondence between quantum
channels and entangled states are made via Stinespring dilation, as is first
proposed Matsumot~et.~al. \cite{MSW}. Due to the Stinespring dilation, 
a CPTP map $\Lambda $ is expressed as the composition of the isometric embedding $U$
followed by the partial trace,%
\begin{equation}
\mathcal{B(K)\overset{U}{\hookrightarrow }B}
(\mathcal{H}_{1}\otimes \mathcal{H}_{2})
\overset{{\rm tr}_{\mathcal{H}_{1}}}{\longrightarrow }
\mathcal{B}(\mathcal{H}_{2}).
\label{stinespring}
\end{equation}
Below, we denote $U{\cal K}$ simply by ${\cal K}$,
so far no confusion arises.
In this correspondence,
\begin{equation}
S\left(\Lambda(\pi )\right)
=E(\pi ).
\label{E-S}
\end{equation}

\section{Yet another additivity conjecture}

Now, we propose a new entanglement quantity,%
\begin{equation}
E_{m}\left( \rho \right) :=\min_{\{p_{i},\pi _{i}\}}\min_{i}E(\pi _{i}),
\label{Emin}
\end{equation}
where $\{p_{i},\pi _{i}\}$ runs all over the ensembles of pure bipartite
states with $\sum_{i}p_{i}\pi _{i}=\rho $. The additivity and the strong
superadditivity of this quantity means, 
\begin{equation}
E_{m}\left( \rho \otimes \rho^{\prime}\right) =E_{m}\left( \rho \right)
+E_{m}\left( \rho \right) ,  \label{addEmin}
\end{equation}
and%
\begin{equation}
E_{m}\left( \rho \right) \geq E_{m}(\mathrm{tr}_{\mathcal{H}^{\prime }}\rho
)+E_{m}(\mathrm{tr}_{\mathcal{H}}\rho ),  \label{supaddEmin}
\end{equation}
respectively. Note that,%
\begin{equation}
E_{m}\left( \rho \right) =\min_{\pi }E\left( \pi \right) ,  \label{Emin-2}
\end{equation}
where $\pi $ runs all over the pure states living in the support of $\rho $.
This expression strongly suggest the close tie between $E_{m}$ and the
output minimum entropy.

\begin{thm}
(main theorem)The followings are equivalent.

\begin{description}
\item[(i)] (\ref{supaddEmin}) for all the pure states.

\item[(ii)] (\ref{supaddEmin}) for all the states.

\item[(iii)] (\ref{addEmin}) for all the states.

\item[(iv)] (\ref{addminent}) for all the quantum channels.

\item[(v)] (\ref{eq:entf:superadd}) for all the states.
\end{description}
\end{thm}

Combining this theorem with the main theorem of \cite{Shor:equivalence}, we
can conclude the additivity of the new entanglement quantity is equivalent
to all the other additivity conjectures.

\begin{pf}
For (iv)$\Leftrightarrow $(v) due to \cite{Shor:equivalence}, it suffices to
show (v)$\Rightarrow $(i)$\Leftrightarrow $(ii)$\Rightarrow $(iii)$%
\Rightarrow $(iv). In the following, let $\rho \in \mathcal{S(H\otimes H}%
^{\prime }).$

\noindent (v)$\Rightarrow$ (i): Let $\rho $ be a pure
state. Then, 
\begin{align*}
E_{m}\left( \rho \right) =E(\rho )=E_{f}(\rho )\geq
 E_{f}(\mathrm{tr}_{\mathcal{H}^{\prime }}\rho )
+E_{f}(\mathrm{tr}_{\mathcal{H}}\rho )\\
\geq E_{m}(\mathrm{tr}_{\mathcal{H}^{\prime }}\rho )
+E_{m}(\mathrm{tr}_{\mathcal{H}}\rho ).
\end{align*}%
(i)$\Rightarrow$ (ii): Let $\pi _{\ast }$ be a pure state living in the
support of $\rho $ with $E_{m}\left( \rho \right) =E\left( \pi _{\ast
}\right) $. Then,%
\begin{align*}
E_{m}\left( \rho \right) =E(\pi _{\ast })
\geq E_{m}(\mathrm{tr}_{\mathcal{H}^{\prime }}\pi _{\ast })
+E_{m}(\mathrm{tr}_{\mathcal{H}}\pi _{\ast })\\
\geq
E_{m}(\mathrm{tr}_{\mathcal{H}^{\prime }}\rho )+E_{m}(\mathrm{tr}_{\mathcal{H}}\rho ),
\end{align*}%
in which the second inequality comes from the assumption, and the third
inequality due to the fact that the support of 
$\mathrm{tr}_{\mathcal{H}^{\prime }}\pi _{\ast }$ 
is a subset of the support of $\mathrm{tr}_{\mathcal{H}^{\prime }}\rho$ .

\noindent (ii)$\Leftarrow$ (i), (ii) $\Rightarrow$ (iii): trivial.

\noindent (iii)$\Rightarrow :$ (iv) 
Let  $\Lambda'$ be a CPTP map from ${\cal B}({\cal K}')$ to
${\cal B}({\cal H}_2')$, and consider isometric embedding 
like (\ref{stinespring}).
Let $\rho$ and $\rho ^{\prime }$ be the
state whose support is $\mathcal{K}$
and $\mathcal{K}^{\prime }$, respectively. 
Then, (\ref{Emin-2}) and (\ref{E-S}) imply, 
\begin{align*}
E_{m}\left( \rho \right) 
&=\min_{\phi\in{\cal K}}
S\left({\rm tr}_{{\cal H}_2}\ketbra{\phi}\right)\\
&=\min_{\phi\in{\cal K}}
S\left(\Lambda(\ketbra{\phi})\right)\\
&= S_{\rm min}(\Lambda'),
\end{align*}
and 
\begin{align*}
E_{m}( \rho' ) 
= S_{\rm min}(\Lambda').
\end{align*}.
In addition, 
for the support of 
$\rho \otimes \rho^{\prime }$ is 
$\mathcal{K}\otimes \mathcal{K}^{\prime }$,
we have,
\begin{equation}  
E_{m}\left( \rho \otimes \rho ^{\prime }\right) 
=S_{\rm min}\left(\Lambda\otimes \Lambda ^{\prime }\right).
\end{equation}
Combining there equations, we have the assertion.
\end{pf}

\section{Properties of $E_{m}$}
For a quantity to be a proper entanglement measure, 
that quantity should be
\begin{description}
\item[(i)] eaqual to $E$ for the pure states.

\item[(ii)] monotone by the application of LOCC.

\item[(iii)] asymptotic continuity.
\end{description}
Our new quantity $E_{m}$ trivially satisfy (i). Also, (ii) is
satisfied, for,letting $\Omega $ be a LOCC operation, and $\pi _{\ast }$ be
the pure state with $E_{m}\left( \pi _{\ast }\right) =E_{m}\left( \rho
\right) $, we have, 
\begin{equation*}
E_{m}\left( \Omega (\rho )\right) \leq E_{m}\left( \Omega (\pi _{\ast
})\right) \leq E_{f}\left( \Omega (\pi _{\ast })\right) \leq E_{f}\left( \pi
_{\ast }\right) =E\left( \pi _{\ast }\right) =E_{m}\left( \rho \right) .
\end{equation*}

However, it is obvious that (iii) cannot be satisfied. 

On the other hand, this quantity satisfies convexity,%
\begin{equation*}
E_{m}\left( p\rho _{1}+\left( 1-p\right) \rho _{2}\right) \leq \min \left\{
E_{m}\left( \rho _{1}\right) ,E_{m}\left( \rho _{2}\right) \right\} \leq
pE_{m}\left( \rho _{1}\right) +\left( 1-p\right) E_{m}\left( \rho
_{2}\right) .
\end{equation*}

\section{Discussions}

Among all the additivity conjectures which are equivalent with each other,
many people are focusing on additivity of the minimum output entropy.
However, in the existing proofs of this additivity conjecture for the
special cases (e.g., \cite{King1, King2, Shor02, MatsumotoYura}), they first
show the strong superadditivity of $E_{m}$ for all the pure states, living
in $\mathcal{K\otimes K}^{\prime }$, 
\begin{equation*}
E\left( \rho \right) 
\geq E_{m}(\mathrm{tr}_{\mathcal{H}^{\prime }}\rho)
+E_{m}(\mathrm{tr}_{\mathcal{H}}\rho ),
\end{equation*}
which naturally leads to the additivity of the minimum output entropy.

Also, in many states for which the additivity or the strong super additivity
of EoF is shown, EoF is equal to $E_{m}$ (e.g.,\cite{VDC, MatsumotoYura}).

Hence, the additivity or the strong superadditivity of $E_{m}$ can be
another good equivalent statement of the additivity conjecture. However, its
operational meaning is hard to find out.

\end{document}